\documentclass[aps,twocolumn,groupedaddress,showpacs]{revtex4}
\usepackage{amsmath,amscd,amssymb,epsfig}

\usepackage{amsmath}
\usepackage{graphicx}
\newcommand{\M}[1]	{\mathbf{#1}}
\renewcommand{\v}[1]	{\roarrow{#1}}

\newcommand{\m}[1]	{\langle\! #1 \!\rangle}

\renewcommand{\t}	{\tau}
\renewcommand{\a}	{\alpha}
\renewcommand{\d}	{\delta}
\newcommand{\w}		{\omega}
\newcommand{\g}		{\gamma}

\begin{document}
\bibliographystyle{apsrev}

\title[]{Markovian embedding of non-Markovian superdiffusion}
\author{Peter Siegle}
\author{Igor Goychuk}
\author{Peter Talkner}
\author{Peter H\"anggi}
\affiliation{Institute of Physics, University of Augsburg,
Universit\"atsstr. 1, D-86135 Augsburg, Germany}

\date{\today}

\begin{abstract} We consider different Markovian embedding schemes
of non-Markovian stochastic  processes that are described by generalized Langevin equations (GLE) and obey thermal detailed balance
under equilibrium conditions. At thermal equilibrium superdiffusive behavior can emerge if the total integral of the memory kernel
vanishes. Such a situation of vanishing static friction is caused by a
super-Ohmic thermal bath. One of the
simplest models of ballistic superdiffusion is determined by a
bi-exponential memory kernel that was proposed by Bao
[J.-D. Bao, J. Stat. Phys. {\bf 114}, 503 (2004)].
We show that this non-Markovian model
has infinitely many different 4-dimensional Markovian
embeddings. Implementing numerically the simplest one, we
demonstrate  that (i)
the presence of a periodic potential with arbitrarily low barriers
changes the asymptotic large time behavior from free  ballistic
superdiffusion into normal diffusion;
(ii) an additional biasing force renders the asymptotic dynamics
superdiffusive again.
The development of
transients  that display a qualitatively different behavior
compared to the true large-time asymptotics presents a general feature of
this non-Markovian dynamics.
These transients though may be extremely long. As a consequence,  they can be even mistaken
as the true asymptotics.
We find that such intermediate
asymptotics exhibit a giant enhancement of superdiffusion in tilted
washboard potentials and it is accompanied by a giant transient
superballistic current growing proportional to $t^{\alpha_{{\rm eff}}}$ with
an exponent
$\alpha_{\rm eff}$
that can exceed the ballistic value of two.

\end{abstract}


\maketitle

\section{Introduction}

The subject of anomalous diffusion has become increasingly popular and
important in the
last years with a number of papers growing faster than linearly in time
with almost 500 papers published last year. There also is a large number of
theoretical models that lead to anomalous diffusion such as
as continuous
time-random walks
\cite{Scher,Shlesinger,Hughes,Bouchaud},
including
Levy flights and Levy walks \cite{Hughes,Metzler},
related
fractional Fokker-Planck equations \cite{MetzlerPRL,Metzler}
and (ordinary) Langevin equations in random subordinated
time \cite{Fogedby,Stanislavsky}, as well
as (ordinary) Langevin equations with additive non-Gaussian Levy white
noises \cite{Ditlevsen,Dubkov}. Moreover nonlinear
Brownian motion with multiplicative Gaussian white noise
\cite{Marksteiner,LutzPRL},  as well as
linear Boltzmann
equation with scattering events being distributed in time according to a
power law distribution \cite{BarkaiSilbey,Friedrich} may display anomalous
diffusion.
This list by far is not
complete. Yet
the quest for minimal and fundamental
physical  models has become ever more important.
One of the
fundamental approaches to anomalous diffusion
\cite{Wang,Morgado,Pottier,WeissBook}
is provided by the
generalized Langevin equation (GLE) \cite{Kubo,Bogolyubov,Zwanzig,HTB90}
with a frictional memory kernel
$\gamma(t)$, reading:
\begin{equation}\label{GLE}
m\ddot x+m\int_{0}^{t}\gamma(t-t')\dot x(t')dt'+
\frac{\partial V(x,t)}{\partial x}
=\zeta(t) \;,
\end{equation}
where $x(t)$ denotes the position of a particle of mass $m$. Here
$\zeta(t)$ is a Gaussian zero-mean fluctuating force that at
temperature $T$ is related to the memory
kernel by the fluctuation-dissipation relation \cite{Kubo}
\begin{equation}\label{FDT} \langle
\zeta(t)\zeta(t')\rangle=k_BTm\gamma(|t-t'|) \;.
\end{equation}
Remarkably, this model can be derived from a Hamiltonian
dynamics of a particle that bilinearly couples with coupling constants $c_i$
to a thermal bath of harmonic oscillators with masses $m_{i}$ and frequencies
$\omega_i$, $H_{B,\rm int}(p_i,q_i,x)=(1/2)\sum_i \{p_i^2/m_i+m_i\omega_i^2
[q_i-c_i x/(m_i\omega_i^2)]^2\}$.
The total effect of the bath oscillators,
which are initially canonically distributed with $H_{B,\rm int}$ 
at temperature $T$ and fixed $x=x(0)$, 
is characterized
by the bath spectral density
\begin{equation}
J(\omega) = \frac{\pi}{2} \sum_i
\frac{c_i^2}{m_{i} \omega_{i}} \delta(\omega-\omega_i).
\label{J}
\end{equation}
It is related to the power spectral density of the fluctuating force
\begin{equation}
  S(\omega)=\int_{-\infty}^{\infty}\langle\zeta(\t)\zeta(0)
  \rangle e^{-i\w\t}d\t
\label{S}
\end{equation}
via $S(\omega)=2k_BT J(\omega)/\omega$ \cite{Kubo,HTB90}.
This in general leads to a non-Markovian process of the particle
dynamics with {\it linear} memory friction and Gaussian fluctuating force.
Moreover, due to the fluctuation-dissipation relation (\ref{FDT}) it is
compatible with thermal equilibrium in confining, time-independent potentials
and encompasses a whole set of physically meaningful models
characterized by different
bath spectral densities $J(\omega)$.

In the absence of any potential the variance of the particle's
position will grow with time. The law according to which the variance
grows characterizes the nature of the resulting diffusion process as
being subdiffusive if the growth of the variance is slower than linear.
This happens if the static friction $\gamma =\int_0^{\infty}\gamma(t')dt'$
diverges. Normal diffusion corresponds to a
linear growth. It occurs if  $\gamma$ is
finite. Finally, if $\gamma$ vanishes the variance grows faster than
linear and one speaks of superdiffusion.
The presence of a nonlinear time-dependent force $f(x,t)=-\partial
V(x,t)/\partial x$ modifies this simple picture in a complicated way
depending on further details of the memory kernel and also on temperature.
The qualitative behavior of the variance of the position is determined
by the variance of the velocity, $\langle\Delta v^2(t)\rangle$. The mean square
displacement of position spreads according to normal diffusion, if the
integral of the velocity variance over all times is finite. On the
other hand if this integral is zero,
then the motion is anti-persistent and
subdiffusive. If this integral diverges the spread of the position
variance is superdiffusive. For free motion in the absence of a
potential these criteria are equivalent to those for the memory
kernel which in general fail in the presence of a potential.
Since general
analytical results are scarce and most likely nonexistent for nonlinear
and time-dependent forcing, the reliability of numerical simulations
has become a key issue.
Numerically tractable models can be obtained by approximating the
given memory kernel by a finite sum of exponential functions.
The
according non-Markovian particle dynamics can then be obtained
as the projection of a
high-dimensional Markovian process onto the phase space of the
particle spanned by the particle's coordinate and momentum $p=m
\dot{x}$.
The dimensionality of the Markovian process is $D=N+2$, where
$N$ is the number of exponentials in the sum approximating the memory
kernel.
The key point is that the corresponding Markovian dynamics can
be propagated locally in time for very long 
time intervals by means of very reliable algorithms with
a well controlled numerical precision. 
Moreover, this way of  thinking allows one to
identify the simplest models for the superdiffusive GLEs with
minimal embedding dimensions $D=3$ and $D=4$. The case $D=4$
corresponds to approximating the memory kernel by a difference of two
exponentials,
\begin{eqnarray}\label{kernel}
\gamma(t)=\gamma_1\exp(-k_1 t)-\gamma_2\exp(-k_2 t),
\end{eqnarray}
such that $\gamma_1/k_1=\gamma_2/k_2$ implying to vanishing static
friction $\gamma$ and  $\gamma_1>\gamma_2$.  The latter
condition amounts to the fact that the memory kernel is proportional
to the autocorrelation function of the fluctuating force $\zeta$ and
hence must be non-negative semidefinite.
This bi-exponential model was  proposed
by Bao \cite{Bao04}. It corresponds to a super-Ohmic spectral density
of thermal bath oscillators, $J(\omega)\propto \omega^3$ at low
frequencies and describes the coupling of a particle to
three-dimensional lattice phonons. It therefore models the
diffusion of an impurity in a crystal.
We will demonstrate that this model can be embedded in infinitely many
ways. In the following we will study one of the simplest embeddings
which is different from those used by Bao  \cite{Bao04,Bao06,Bao05a}.
In the absence of any force $f(x,t)$
the spreading of the particle's position
distribution is ballistic,
$\langle \Delta x^2(t)\rangle = D_{2} t^2$, and hence super-diffusive. 
Here and in the following, the expectation
$\langle...\rangle$ refers to an ensemble 
average with respect to the fluctuating
force $\zeta(t)$ and an initial distribution of position and momentum,
$x(0)$ and $p(0)$, respectively.
The velocity process is
{\it non-ergodic} \cite{Bao05}. As a consequence,
the ballistic superdiffusion coefficient $D_{2}$ turns out to depend
on the initial
velocity distribution.
However, as we shall see below this
non-ergodic feature disappears in periodic potentials. Moreover, our
numerics reveals that the
ballistic diffusion in tilted periodic potentials does neither depend
on the initial velocity distribution, nor on the initially non-equilibrium
noise preparation. For this reason, we presume that the velocity
process of the particle then
becomes ergodic.

The minimal, three-dimensional Markovian embedding of the GLE
superdiffusion is achieved in the limit $\gamma_1\to\infty$,
$k_1\to\infty$, so that $\gamma_0=\gamma_1/k_1={\rm
const}=\gamma_2/k_2$. In this limit, the first exponential becomes
a delta-function $2\gamma_0\delta(t)$.
This case will be studied elsewhere.

Unfortunately, a non-Markovian Fokker-Planck-type
equation (NMFPE) that corresponds to a GLE with a general, nontrivial
potential is not known in spite of many years of search.
The only exceptions are provided by (strictly) linear and parabolic
potentials, where the corresponding NMFPEs were derived by Adelman \cite{Adelman} and H\"anggi \cite{hanggi,hanggi2,hanggi3}
for stable non-Markovian Brownian motion  and, as well, for unstable
non-Markovian dynamics \cite{hanggiUNM}; i.e. the Kramers problem  of escape over a 
parabolic barrier \cite{HTB90}. Using the fact
that  $[x(t),\dot x = v(t)]$ is a two-component
Gaussian process, which is obtained by a linear integral transformation
of the Gaussian noise process  $\zeta(t)$ the resulting NMFPE assumes the form of a time-dependent FPE. 
This FPE-structure of time-evolution for the single-time event probability of the non-Markovian process 
should then not be mistaken as an effective
Markovian dynamics \cite{hanggi,hanggi2,hanggi3}. Notwithstanding these  exceptional known cases cases of  
non-Markovian Gaussian dynamics, this lack of a generally closed NMFPE for nonlinear forces lends
even more importance to the Markovian embedding approach.

This paper is structured as follows. In Sec. II and Appendix I
we detail the
Markovian embedding procedure in a slightly more general way
than has been used so far. The general results are illustrated with
two different embeddings of one and the same superdiffusive GLE dynamics.
In Sec. III, we present and discuss the results of stochastic simulations of superdiffusion under a constant bias and in a washboard potential for one of these embeddings.
The issue of ergodicity of mean square displacement is discussed in Sec. IV.
Conclusions are drawn in Sec. V.

\section{Method}

The idea that we pursue here
is to represent the non-Markovian stochastic dynamics of a single
particle with
$(x,p)$ phase space
as a projection of a multi-dimensional Markovian dynamics. It
is well known that any GLE can be derived from the (Markovian)
Hamiltonian dynamics  of a particle coupled to a thermal bath of
harmonic oscillators. This Hamiltonian embedding though requires a large
number of auxiliary degrees of freedom representing the thermal bath.
Here we look for an embedding with a minimal number $N$ of auxiliary
variables which all together constitute a continuous Markovian process.
A low embedding dimension is
crucial for running numerical simulations
which can become extensively time-consuming for a large $N$.

We first rewrite the GLE (\ref{GLE}) in terms of the phase space
coordinates $x$ and $p = m \dot{x}$
as
\begin{align}\label{eqn:init}
 \dot x(t)=&\frac{1}{m} p(t) \notag\\
 \dot p(t)=& f(x,t)-
 \int_{0}^{t} \g(t-t^\prime) p(t^\prime)dt^\prime+ \zeta(t)\;.
\end{align}
The embedding involves a yet to be determined number $N$ of
auxiliary dynamical variables collected into a vector $\vec u(t)$
in terms of which the dynamics takes the following general form
\begin{align}\label{eqn:inan}
 \dot x(t)=&\frac{1}{m}p(t) \notag\\
 \dot p(t)=&f(x,t)+ \v g^\mathrm{T}
 \v u(t)\notag\\
 \dot{\v u}(t)=&-p(t)\v r -\M{A} \v u(t)+ \M{C} \v \xi(t)\;,
\end{align}
where $\v g$ and $\v r$ denote constant vectors of dimension
$N$ and $\M{A}$, and $\M{C}$ are constant $N\times N$ matrices. The
upper index $T$ denotes the transpose of a vector or a matrix.
Further,
$\v\xi(t)$ is a vector of uncorrelated Gaussian white noises,
\begin{align}\label{xicorr}
 \langle \xi_i (t)\xi_j (t^\prime)\rangle =\d(t-t^\prime)\d_{ij}\;
\end{align}
with $N$ components. Integrating the equation for the auxiliary vector
$\v u(t)$ and
substituting the result
in the equation for the momentum $p(t)$ one recovers the original GLE
(\ref{eqn:init}) only
under special conditions,
see Appendix A for details
of the derivation. First, the memory kernel $\gamma(t)$
must satisfy
\begin{eqnarray}\label{first}
\g(t) =& \v g^\mathrm{T} e^{-\M{A}t}\v r\;.
\end{eqnarray}
Since the right hand side can in general be represented as a sum of
$N$ exponential functions $\exp(- \lambda_{i}t)$, $i=1\ldots N$ with
the eigenvalues $\lambda_{i}$ of the matrix $\M{A}$
the embedding can be exact only if the memory kernel is of the same
type \cite{Jordan}. But also other memory kernels such as
algebraically decaying functions can be  approximated by a finite
sum of exponential functions, even with a relatively small extra
dimension $N$, and hence are amenable to Markovian embedding.\\
Furthermore, 
the fluctuation-dissipation relation (\ref{FDT}) imposes restrictions on the matrices $\M{A}$, $\M{C}$, and the
vectors $\v g$ and $\v r$. These restrictions are met if the embedding
parameters satisfy the following two relations:
\begin{align}\label{Gtor}
\M{G}\v g &= m k_B T \v r \\
\M{C}\M{C}^\mathrm{T}& = \M{A}\M{G}+\M{G}\M{A}^\mathrm{T} \;\label{eqn:cond},
\end{align}
which defines the constant $N\times N$ matrix $\M{G}$.\\
However, for arbitrary initial values
of the auxiliary variables $\v u(0)$, the fluctuation-dissipation relation
will be obeyed only asymptotically. This means that the noise $\zeta(t)$
in  Eq.~(\ref{eqn:init}) is initially nonstationary and becomes only gradually
stationary in the course of time, see Appendix A, Eq.~(\ref{eqn:statnoise}). 
In order to guarantee the Gaussian nature of the random
force $\zeta(t)$ the vector $\v u(0)$ must also be Gaussian
distributed, see Eq.(\ref{eqn:z}).
Because  the vector $\v u(0)$ is
independent of the vector of Gaussian white noises $\v \xi(t)$
and its first moment must vanish, it is sufficient to specify its
covariance matrix
\begin{eqnarray}
\langle \v u(0)  \otimes \v u^\mathrm{T}(0)\rangle = \M{G}\;.
\label{eqn:UG}
\end{eqnarray}
It must coincide with $\M{G}$ in Eq. (\ref{Gtor}), in order to have 
the fluctuation-dissipation relation (\ref{FDT}) obeyed for all times,
 see Eq. (\ref{eqn:statnoise}).

Yet the conditions (\ref{first}), (\ref{Gtor}) and (\ref{eqn:cond}) do not
uniquely determine the enlarged process and actually leave room for an
infinite variety of different processes leading upon reduction to the
same generalized Langevin equation. Since some of the enlarged
processes allow faster
and more reliable numerical simulations than others there is a great
interest in identifying computationally optimal embeddings.
We further note that the relations (\ref{first}), (\ref{Gtor}) and
(\ref{eqn:cond}) are sufficient but not necessary conditions. The
resulting embedding is
more general than previous ones such as those proposed in
Refs.~\cite{Kupferman,Marchesoni} which assume $\v r = \v g$.

Before discussing a particular example we would like to emphasize that the stationarity of the fluctuating force and the fluctuation
dissipation relation~(\ref{FDT}) are exactly
implemented.

\subsection{ Minimal model}

We now consider the simple class of models specified by the
bi-exponential memory kernel (\ref{kernel}).
Under the condition of vanishing static friction, i.e.
 $\gamma_1k_2=\gamma_2k_1$, this memory kernel is specified by three
independent
parameters that can be written as
\begin{align}
\kappa^2= \g_1-\g_2,\quad \nu = k_1+k_2 \quad\textrm{and}\quad \w_0^2= k_1k_2\;.
\end{align}
Note that $\kappa$ is always a real parameter due to the positivity
constraint $\gamma_{1}>\gamma_{2}$. In terms of this parameterization
the Laplace transform $\hat \g (s)$ of the memory kernel becomes
\begin{equation}
\hat \g (s)=\int_0^\infty e^{-st} \g(t) dt =\frac{\kappa^2
  s}{s^2+\nu s+\w_0^2}\;.
\label{gs}
\end{equation}
One can now easily calculate the power spectral density, see Eq.~(\ref{S}), by connecting it to the friction kernel via the fluctuation-dissipation relation, see Eq.~(\ref{FDT}):
\begin{equation}
\begin{split}
S(\w)&=2 k_B T m\int_0^\infty \g(t)\cos(\w t)dt \\
&= 2k_BT m\operatorname{Re}\big(\hat\g(i\w)\big)\\
&=
\frac{2 k_B T m \kappa^2\nu\w^2}{(\w^2-\w_0^2)^2+\nu^2\w^2}\;.
\end{split}
\end{equation}
It is interesting to note that the power spectral density has a
maximum at the frequency $\w_{0}$.

There are several possibilities to realize Markovian embedding  of
this non-Markovian model. In the following we shall discuss two of them.

\subsubsection{First embedding }

A simple embedding is obtained by choosing:
\begin{equation}\label{eqn:firstembedding}
\begin{split}
 \M{A}&=\left(\begin{array}{cc} \nu & \w_0 \\ -\w_0 & 0
     \\ \end{array}\right) \\ \v g^{T} &= \v r^{T} =
 \left(\kappa ,\;0\right) \\
 \M{C}&=\sqrt{2 m k_B T \nu}\left(\begin{array}{cc} 1 & 0 \\ 0 & 0
     \\ \end{array}\right) \\
\langle u_i(0) & u_j(0)\rangle =m k_B T\delta_{ij} \;.
\end{split}
\end{equation}
This choice leads to the following equations:
\begin{equation}\label{eqn:4dim}
\begin{split}
 \dot x(t)=&\frac{1}{m}p(t) \\
 \dot p(t)=&-\frac{\partial}{\partial x} V(x,t) + \kappa u_1(t) \\
 \dot u_1(t)=&-\kappa p(t) - \nu u_1(t) - \w_0 u_2(t)\\
&+ \sqrt{2 m k_B T \nu} \xi(t) \\
 \dot u_2(t)=& \w_0 u_1(t) \;,
\end{split}
\end{equation}
where $\xi(t)$ is scalar Gaussian white noise.
Moreover, this embedding also allows for complex parameters $k_1=k_2^*$,
providing the possibility to model oscillating
real valued kernels
\begin{eqnarray}
\gamma(t)&= &\kappa^2 e^{-\nu t/2} [ \cos(t\sqrt{\omega_0^2-\nu^2/4}) \\
& - &\frac{\nu}{\sqrt{4\omega_0^2-\nu^2}}
\sin(t\sqrt{\omega_0^2-\nu^2/4})]\;. \nonumber
\end{eqnarray}
This model corresponds to
sharply peaked power spectral density $S(\omega)$, and bath spectral
density $J(\omega)$.

\subsubsection{Second embedding}
An alternative way is to start with a diagonal matrix $\M{A}$. It
would be tempting to also choose diagonal matrices $\M{C}$ and
$\M{G}$. This choice though always yields a linear combination of
exponential functions with positive coefficients for the memory kernel
(\ref{kernel}) \cite{Goychuk09} and hence does not allow vanishing
static friction. However,  this goal can be achieved by means of the
following choice
of parameters involving non-diagonal matrices $\mathbf{C}$ and $\mathbf{G}$
\begin{equation}\label{eqn:secondembedding}
\begin{split}
 \M{A}&=\left(\begin{array}{cc} k_1 & 0 \\ 0 & k_2
     \\ \end{array}\right)  \\
 g_{1,2} & = \sqrt{\gamma_{1,2}\frac{k_1+k_2}{k_1-k_2}}\\
 r_{1,2} & =
 \sqrt{\gamma_{1,2}\frac{k_1-k_2}{k_1+k_2}}\\
 \M{C}& =\sqrt{2 m k_B T }\left(\begin{array}{cc} \sqrt{k_1} & 0 \\
     -\sqrt{k_2} & 0 \\ \end{array}\right)   \\
\M{G}
& =m k_B T  \left(\begin{array}{cc} 1 & -c \\ -c & 1 \\ \end{array}\right)\;,
\end{split}
\end{equation}
where $0<c=2\sqrt{k_1k_2}/(k_1+k_2)<1$ is the correlation
coefficient of the covariance matrix $\M{G}$
and $k_1=k_2(\gamma_1/\gamma_2)>k_2$.  This choice is similar to the
one in \cite{Bao04}.
We note that the second embedding requires that the parameters $k_{1}$
and $k_{2}$ must be real. Therefore it is not possible to model an
oscillating kernel by this method.
Our first embedding in Eq. (\ref{eqn:4dim}) is, however, simpler and
numerically
more convenient since its numerical simulation requires less operations.
For instance, only one stochastic variable has to be generated in the
first embedding scheme, see Eq. (\ref{eqn:firstembedding}), in contrary
to two, needed in the second embedding scheme, see. Eq.(\ref{eqn:secondembedding}).
All this makes our first embedding scheme preferential.

\subsection{Dimensionless units}

For further studies, we transform Eq.~(\ref{eqn:4dim})
into dimensionless units by scaling momentum in terms of thermal momentum
$p_T=\sqrt{m k_BT}$, expressing the distance in terms of a typical
length scale some arbitrary length $x_0$,
which becomes the spatial period for periodic potentials (see below) and time in
units of $\tau_0=x_0/v_T$. The auxiliary variables $u_{1}$ and $u_2$ are scaled in
units of $u_{0} = m x_{0}/(\kappa \tau^{2}_{0})$.
The energy  is scaled in units
of $k_BT$. This yields the equations of motion
\begin{align}\label{eqn:4dimless}
 \dot{\tilde x}=&\tilde v \notag\\
 \dot{\tilde v}=&-\frac{\partial}{\partial \tilde x} \tilde V(\tilde x,\tilde t) + \tilde u_1 \notag\\
 \dot{\tilde u}_1=&-\tilde \kappa^2 \tilde v - \tilde\nu \tilde u_1 - \tilde \w_0 \tilde u_2 + \tilde \kappa\sqrt{2\tilde\nu} \xi \notag\\
 \dot{\tilde u}_2=& \tilde \w_0 \tilde u_1 \;,
\end{align}
which were used in our simulations.
Here, $\tilde \kappa=\kappa\tau_0$, $\tilde \omega_0=\omega_0\tau_0$,
$\tilde \nu=\nu\tau_0$.
All results in the following figures are given in these dimensionless units.

\section{Results}

The numerical results presented below were obtained using the standard
stochastic Euler  method \cite{Gard}. A Mersenne Twister pseudo random
number generator was used to produce uniformly distributed random
numbers which were transformed into Gaussian variables using
Box-Muller algorithm \cite{NumRec}.  Typically, an ensemble of
$n=10^4$ particles (or trajectories) was propagated in time with a
fixed time step between $\Delta t=10^{-4}$ and $10^{-5}$ in most
simulations to achieve (weak) convergence of  ensemble averaged
results. The use of double precision thus cannot be avoided and
reliable numerics are very time consuming. 
All the particles were
initially localized  at $x(0)=0$ with the initial velocities sampled
from some probability distribution. In most simulations we assumed
this distribution to be sharply peaked at zero and ascribed zero 
initial velocities to all the particles, although the thermal Maxwellian
distribution was also used. The auxiliary variables $u_i(0)$ were
(mostly) sampled from the corresponding Gaussian distributions to
achieve at the exact equivalence  of the simulated Markovian dynamics to
that of GLE, as described in the Section II. Method. Sometimes, we used
also a different initial distribution of $u_i(0)$ (all equal zero) in
order to clarify the influence of the initially non-equilibrium noise
preparation on the stochastic dynamics. In all cases, we denote the
corresponding ensemble averages as $\langle...\rangle$ and specify
the initial distributions if not obvious. 

Of central interest are the first moment $\langle \Delta x(t)
\rangle$ and the variance $\langle\Delta x^2(t)\rangle$ of the displacement
\begin{equation}
\Delta x(t) = x(t) - x(0) = \int_{0}^{t} dt' v(t').
\label{Dx}
\end{equation}
Accordingly we have
\begin{equation}
\langle \Delta x(t)\rangle = \int_{0}^{t} dt' \langle v(t') \rangle
\label{EDx}
\end{equation}
and
\begin{equation}
\begin{split}
\langle\Delta x^{2}(t)\rangle &= \langle(\Delta x(t)- \langle \Delta x(t)
\rangle)^{2} \rangle\\
& = \int_{0}^{t} \int_{0}^{t} dt_{1} dt_{2} C_{v}(t_{1},t_{2})\;,
\end{split}
\label{sigma}
\end{equation}
where $C_{v}(t_{1},t_{2})$ denotes the velocity fluctuation
autocorrelation function
\begin{equation}
C_{v}(t_{1},t_{2})=
\langle(v(t_{1})- \langle v(t_{1})\rangle)(v(t_{2})- \langle
v(t_{2})\rangle) \rangle \;.
\label{Cv}
\end{equation}
These quantities were estimated on the basis of averages over the
ensemble of simulated particle trajectories.

Of particular interest will turn out the question under which
conditions the process
of velocity fluctuations defined as the deviation of velocity from its
mean value constitutes an ergodic process \cite{Deng,Papoulis}.
The definition and main properties of an ergodic process are collected
in Appendix~\ref{B}.

\subsection{Superdiffusion in presence of a constant bias}

First, we consider  the Langevin dynamics (\ref{GLE}) with an arbitrary memory
kernel  under a constant biasing force $F$, i.e with $V(x,t)
=-F x$. This special biased problem is analytically solvable,
cf. \cite{Wang,Morgado, Pottier,Kupferman}, and therefore provides a suitable
test of our
numerical simulations.
The mean square displacement in this case does not depend on the external bias $F$  and is given by:
$\langle \Delta x^2(t)\rangle =
\langle x^2(t)\rangle-\langle x(t)\rangle^2 $ becomes
\begin{eqnarray}\label{exact}
\langle \Delta x^2(t)\rangle  =  2v_T^2\int_0^t H(t')dt'
+ \left [\langle v^2(0)\rangle-v_T^2\right] H^2(t),
\end{eqnarray}
where we denote the thermal average  of initial velocities by
 $ v_T^2 \equiv \langle v^2(0)\rangle_T =k_{B}T/m$ and
\begin{equation}
H(t)=\int_0^{t} K_v(\tau) d\tau
\end{equation}
is the integral of the (normalized) equilibrium autocorrelation function
of the velocity fluctuations which is defined as
\begin{equation}
K_v(\tau)=C_{v}(\tau,0)/v^{2}_{T}\; .
\label{KvC}
\end{equation}
It has the Laplace-transform
\begin{equation}\label{Kv}
\hat K_v(s) = \frac{1}{s+\hat \gamma(s)}\;.
\end{equation}
We note that the velocity fluctuations present a wide sense ergodic
process if and only if
the time average of $K_v(t)$ vanishes, i.e.:
\begin{equation}
\lim_{t\to \infty} \frac{1}{t} H(t) =0,
\end{equation}
see also Appendix~\ref{B}.
For the mean displacement one obtains
\begin{equation}
\langle \Delta x(t)\rangle = \frac{F}{m} \int_0^t H(t')dt'.
\end{equation}
If one chooses thermally distributed initial velocities, $\langle
v^2(0)\rangle=v_T^2$, then the first and second moments of the
displacement are connected by
the fluctuation-dissipation theorem (FDT)
\begin{equation}\label{FDT2}
\langle \Delta x(t)\rangle =\frac{F}{2k_B T}\langle \Delta x^2(t) \rangle ,
\end{equation}
for any memory kernel. Notice that the mass of the particle is not involved
in Eq. (\ref{FDT2}).
Provided that the velocity process is ergodic in the wide sense the
second term on the right hand side of Eq. (\ref{exact}) can be neglected
compared to the first term if time goes to infinity. Hence, for an
ergodic velocity process the spreading of the particle position
becomes independent of the initial velocity distribution. In contrast,
for a nonergodic process the second term can become
comparable in magnitude or even
dominant for large times. Then the influence of the
initial velocity distribution on the second moment of the position
survives. This actually happens if the Laplace
transform of the
memory kernel, $\hat{\gamma}(s)$ approaches zero for $s \to 0$
proportionally to $s$ or faster. In this case, the FDT (\ref{FDT2})
is not valid for $\langle v^2(0)\rangle \neq v_T^2$,
even asymptotically.

As an example we consider the minimal model (\ref{kernel}). Its
Laplace transform indeed vanishes linearly with $s \to 0$, see
Eq.~(\ref{gs}).
For the Laplace transform of the velocity correlation coefficient one
obtains from Eq.~(\ref{Kv})
\begin{equation}
\hat{K}_{v}(s) =\frac{s^{2} +\nu s + \w_{0}^{2}}{s(s^{2}+\nu s
  +w_{0}^{2} +\kappa^{2})}\; .
\label{Ksmm}
\end{equation}
Inverting the Laplace transform, one obtains for the velocity
correlation coefficient
\begin{eqnarray}\label{vel-corr}
K_v(\tau)&=
&\frac{\omega_0^2}{\omega_0^2+\kappa^2}+\frac{\kappa^2}{\omega_0^2+\kappa^2}e^{-\nu   \tau/2}  \\
&\times& \Big[ \cosh(\frac{\sqrt{\nu^2-4\omega_0^2-4\kappa^2}}{2}\tau)
\nonumber \\
& + &  \frac{\nu}{\sqrt{\nu^2-4\omega_0^2-4\kappa^2}}
\sinh(\frac{\sqrt{\nu^2-4\omega_0^2-4\kappa^2}}{2}\tau)\Big] \nonumber \;.
\end{eqnarray}
Note that with $\lim_{\tau \to \infty} K_{v}(\tau) =\w_{0}^{2}/(w_{0}^{2}
+\kappa^{2})$ the equilibrium autocorrelation function of velocity fluctuations as well as its time average
attain positive values. This confirms that the velocity process of
the minimal model is nonergodic in the case of linear potentials.

\begin{figure}
	\includegraphics [width=\linewidth]{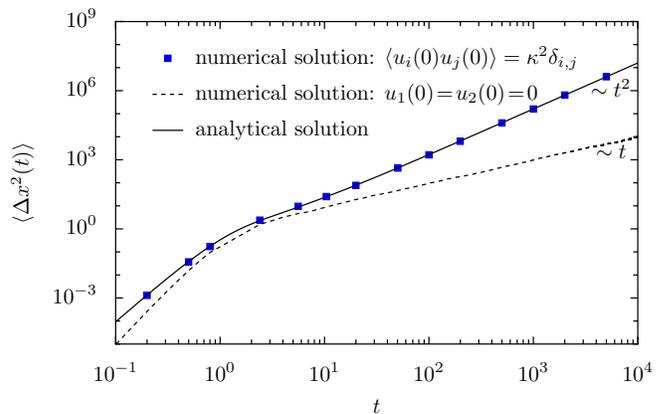}
	\caption{(Color online) The mean square displacement of the position (being  independent of the bias $F$) as a function of time
          changes from a $t^{4}$ law at small times to a ballistic
          $t^{2}$ law. Comparison of the analytical solution according
          to Eq.~(\ref{exact}) (solid line) and results from
          numerical simulations
          of the first embedding (square symbols) exhibit good
          agreement.
          A strongly deviating result is obtained if the auxiliary
          variables of the first embedding initially assume vanishing
          values (dashed line). The other parameters $\kappa=2$,
          $\nu = 3$, $\w_{0}=1$ and $v(0)=0 $ are the same for the
          displayed curves. The estimate of the mean square
          displacement is obtained by an average over an
	  ensemble of $10^{4}$
          simulated trajectories.
}
\label{Fig1}
\end{figure}

The mean square displacement of the position
can be exactly evaluated by means of Eq.~(\ref{exact}). We refrain from
presenting the resulting lengthy expression and only compare the such
derived exact result with the
mean square displacement obtained from a simulation of the Markovian
model via the first embedding for a particular set of
parameters, see Fig.~\ref{Fig1}. The agreement between the analytical
result and the simulation is very good.

For short times the spreading of the mean square displacement of the
position becomes
\begin{eqnarray}
\langle \Delta x^2(t)\rangle & = & \langle v^2(0)\rangle t^2 \nonumber \\
&+ & \kappa^2
\left [ 3 v_T^2-4\langle v^2(0)\rangle \right] t^4/12 + O(t^5).
\end{eqnarray}
For a strictly vanishing initial velocity the contribution
proportional to $t^{2}$ disappears and the diffusion initially becomes
super-ballistic with
$\langle \Delta x^2(t)\rangle\propto t^4$,
see Fig. \ref{Fig1}.
Otherwise, the diffusion initially is  ballistic.
For large times  ballistic diffusion results
with $\langle \Delta x^2(t)\rangle \sim D_2 t^2$. Due to the
nonergodicity of the velocity process
the ballistic superdiffusion coefficient
$D_2$ depends on the initial distribution of velocities,
\begin{eqnarray}\label{Kcoeff}
D_2 &=& v_T^2\frac{\omega_0^2}{\omega_0^2+\kappa^2} \nonumber \\
&\times &\left [ 1+ \left(\frac{\langle v^2(0)\rangle}{v_T^2} -1 \right)
 \frac{\omega_0^2}{\omega_0^2+\kappa^2}  \right] \;.
\end{eqnarray}

Fig.~\ref{Fig1} also displays simulation results of the first
embedding for vanishing  initial values
of the auxiliary variable, i.e. $u_{1}(0)= u_{2}(0)=0$, and also
for initial
values from a Gaussian distribution with variance $\langle u_{i}(0)
u_{j}(0) \rangle = \kappa^{2} \d_{i,j}$. We recall that the latter
choice guarantees that the fluctuating forces are stationary and
that they satisfy the fluctuation-dissipation relation (\ref{FDT}).
The mean square displacement resulting from zero initial auxiliary
variables  is remarkably different from
that with the correct Gaussian distributed initial auxiliary variables
not only at short times but also for large times where it approaches
normal instead of ballistic diffusion. The strong influence of the
initial conditions even at large times is another consequence of the
non-ergodicity of the velocity. In contrast, for an ergodic velocity process the
long time behavior of the position mean square displacement has lost
any memory on initial conditions.

\subsection{Superdiffusion in a washboard potential}

\begin{figure}[htbp]
	\includegraphics [width=\linewidth]{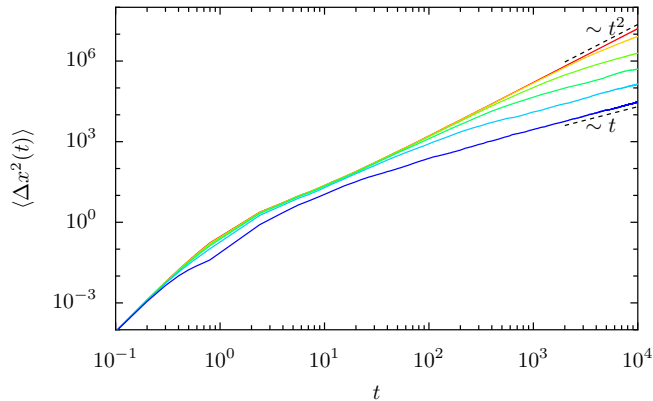}
	\caption{(Color online) Unbiased superdiffusion in a washboard potential. The mean square displacement of a particle that
          spreads ballistically in the absence of a bias $F$, see
          Fig.~\ref{Fig1}, eventually changes its behavior from
          superdiffusive to
          normal diffusive
          behavior under the influence of a
          periodic potential of strength $V_0$.
	  From bottom right hand side to top we use $V_0=0.5,0.2,0.1,0.05,0.02,0$.
	  The time of the turn-over shifts to later times with decreasing potential strength.
	  For the simulation of the displayed
          data the first embedding was used with $\kappa=2$, $\nu=3$,
          $\w_{0}=1$ and $v(0)=0$.
}
\label{Fig2}
\end{figure}

Next we consider the diffusion in a periodic washboard potential
$V(x,t)=-V_0\cos(2\pi x/x_0)$ of spacial period $x_0$.  
Here no analytical results are
available, instead we performed numerical simulations of the first
embedding.
In Fig.~\ref{Fig2} we compare the simulated mean square displacement
as a function of time for different heights of the potential barriers
separating neighboring periods of the potential.
After a short
initial period  of fast growth the diffusion turns over in an
intermediate ballistic behavior which eventually changes into normal
diffusion. As far as one can say from the numerical simulations of
finite duration, normal diffusion always determines the asymptotic behavior.
The onset time of normal diffusion though crucially depends on the magnitude
of the potential barrier $2 V_{0}$. The larger this barrier is the
earlier normal diffusion sets in. On the other hand for small barriers
the ballistic regime extends over a large time before the asymptotic
normal diffusion takes over.

\subsection{Superdiffusion in a biased washboard potential}

\begin{figure}[htbp]
	\includegraphics [width=\linewidth]{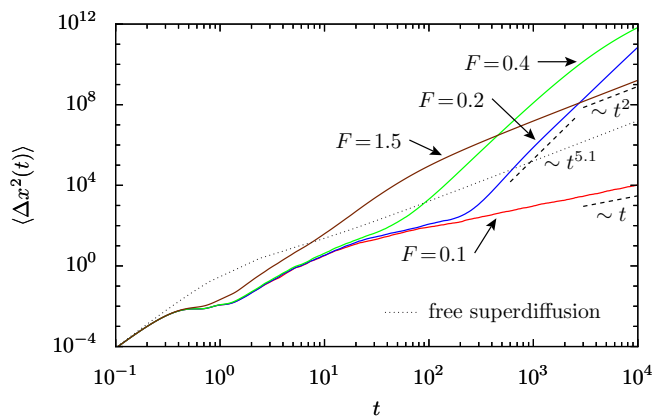}
	\caption{(Color online) Biased superdiffusion.
	After diffusion has become normal in the presence of a periodic potential, see Fig.~\ref{Fig2}, it is again changing to ballistic diffusion under the influence of an additional finite bias $F$.
	Long transients exhibiting  hyper-diffusion emerge before the ballistic diffusion regime is approached.
	A fixed barrier height $V_0=1$ was used for the simulations and the tilt $F$ is variable. The other parameters are again $\kappa=2$, $\nu=3$, $\w_{0}=1$ and $v(0)=0$.
}
\label{Fig3}
\end{figure}
\begin{figure}[htbp]
	\includegraphics [width=\linewidth]{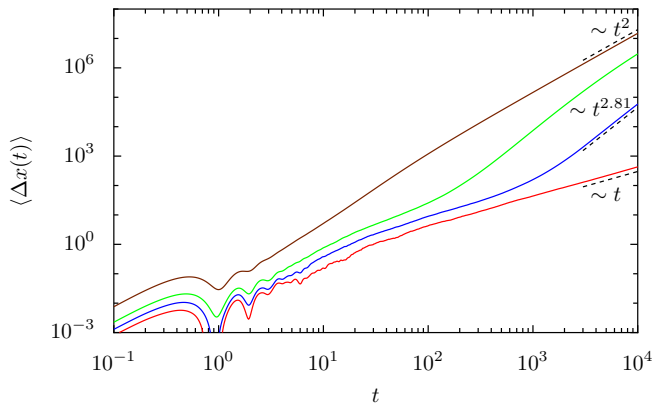}
	\caption{(Color online) Anomalous drift behavior.
	A finite bias $F$ induces  anomalous drift.
	From bottom right hand side to top we use $F=0.1,0.2,0.4,1.5$.
	Ballistic currents appear asymptotically $\langle \Delta x(t)\rangle \sim t^2$.
	Like in Fig. \ref{Fig3} transient regimes appear with enhanced particle transport stronger than ballistic.
	The used parameters are: $V_0=1$, $\kappa=2$, $\nu=3$, $\w_{0}=1$ and $v(0)=0$.}
\label{Fig4}
\end{figure}

The modification of the dynamics by a tilt of the washboard potential,
$V(x,t)=-V_0\cos(2\pi x/x_0)-Fx$,
provides an intriguing question. In particular one may ask whether the
spreading will again become
superdiffusive and whether a supercurrent will emerge that steadily
grows with time? The numerical simulations displayed in
Figs.~\ref{Fig3} and \ref{Fig4} indicate that the answer to both
questions is yes. Both the mean square displacement as well as the
average displacement become proportional to a ballistic law
$t^{2}$. The time at which this presumably asymptotic
behavior sets in becomes increasingly larger the smaller the bias
force $F$ is. For the stronger forces $F=0.7, 1.5$ the ballistic
regime has settled within the total time of $t=10^{4}$ which requires
a week of computational time on a Pentium PC with 3 GHz
tact-frequency.
For small forces
$F=0.1, 0.2$ an approach to a ballistic behavior is not yet visible.
We expect it to occur at a later time.

Another interesting feature is the occurrence of very long
superdiffusive transient episodes with a mean square displacement
growing faster than ballistic as $t^{\alpha_\text{eff}}$ with an exponent
$\a_{\text{eff}}>2$ up to approximately $5$. We call these episodes
``hyper-diffusive''. Their occurrence depends on the dimensionless
barrier height
$V_{0}/k_{B} T$ and the biasing force $F_{0}$. With a larger barrier
the total transient time before the asymptotic ballistic behavior sets
in becomes larger. For small biasing forces after a short initial period
first a regime of normal diffusion is observed which turns over into
the hyper-diffusive regime at a time that is the later the smaller the biasing
force is.
For example, for $V_{0}/k_{B} T =1$ and $F=0.2$ the normal diffusion
regime extends approximately over one decade from $t=10$ to $t=100$,
and then rapidly turns around $t=2\times10^{2}$ into hyper-diffusion with $\alpha_{\rm eff}
\approx 5.1$, cf. Fig.~\ref{Fig3}. This behavior continues until the
end of the simulation at $t=10^{4}$. Until then the  root mean square
displacement increases by an amount of $10^{3}-10^{4}$ periods of
length $x_{0}$.
The turnover to the expected
ballistic diffusion
can only be observed if the biasing force is larger, but then also the
normal diffusion regime disappears.

A similar effect of hyper-diffusive motion was reported by
L\"u and Bao \cite{Lu} for a Brownian particle moving in a biased periodic
potential under the influence of a super-Ohmic model with a spectral
density $J(\w) \propto \w^{1.5}$ for
$\w \to 0$. The question whether the hyper-diffusion observed in
Ref.~\cite{Lu} is indeed asymptotic or whether it is also a transient
phenomenon must still be clarified.

From the different curves displayed in Fig.~\ref{Fig3} one can
infer that for large times the mean square displacement grows the faster
the smaller the biasing forces is, in other words, the ballistic
diffusion constant increases with decreasing biasing force and, in
particular, is larger than the ballistic diffusion constant of free
motion reached for $Fx_0\gg V_0$.
This phenomenon is akin to the effect of giant enhancement
of normal diffusion in periodic potentials   \cite{giant,giant2}.

The mean displacement $\langle \Delta x(t) \rangle$ exhibits a
qualitatively similar behavior as the mean square displacement. After
a first transient period whose nature strongly depends on the initial
velocity distribution
a monotonous growth sets in that changes from linear to quadratic,
possibly interrupted by an episode of rapid growth proportional to
$t^{\beta}$ with $\beta >2$.
cf. Fig. \ref{Fig4}.
The transitions between the different regimes occur at the same times at
which the mean square displacement changes from normal diffusion
into the hyper-diffusion and finally to ballistic diffusion.
The exponent $\beta$ though is much smaller than the hyper-diffusive
exponent  $\alpha_{\rm eff}$. This indicates that the transport in
this intermediate regime is strongly erratic.
While both periods of normal and ballistic diffusion can be
characterized by a time-independent Peclet number
${\rm Pe}= x_0 \langle \Delta x(t)\rangle/\langle \Delta x^2(t)\rangle$
\cite{Lindner} the difference of the exponents $\a_{\rm eff}$ and
$\beta$ does not allow the
definition of a Peclet number in the hyper-diffusive regime.
However, both in the normal and the asymptotic ballistic regime a
time-independent Peclet number can be defined.
For $F=0.1$ the FDT (\ref{FDT2}) holds with a good
accuracy and ${\rm Pe}\approx F x_0/(2 k_BT)$
in the normal diffusion transport regime, see Figs. \ref{Fig3}, \ref{Fig4}.
Beyond the linear response regime, the FDT (\ref{FDT2}) is generally
violated. Such a
wealth of different transport regimes with normal
and anomalous features, revealed by a simple  model is really
surprising.

\section{Ergodicity}

\begin{figure}[htbp]
	\includegraphics [width=\linewidth]{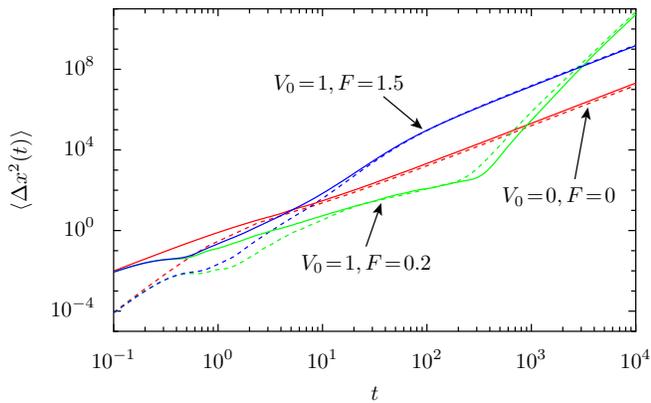}
	\caption{(Color online) Mean square displacement of position for different initially distributed velocities.
	The solid lines mark thermally distributed initial velocities $\langle v^2(0)\rangle=v^2_T$,
	whereas dashed lines mark initially zero velocity $v(0)=0$.
	The differences in time evolution vanish in the asymptotic long time limit in the presence of a biased periodic washboard potential with potential height $V_0$ and bias $F$, implying wide sense ergodicity for the mean square displacement.
	However, in case of free ballistic diffusion, see lines labeled by $V_0=0,F=0$,
	a constant deviation remains according to Eq. (\ref{Kcoeff}).
	Parameters are chosen as $\kappa=2$, $\nu=3$ and $\w_{0}=1$.}
\label{Fig5}
\end{figure}

\begin{figure}[htbp]
	\includegraphics [width=\linewidth]{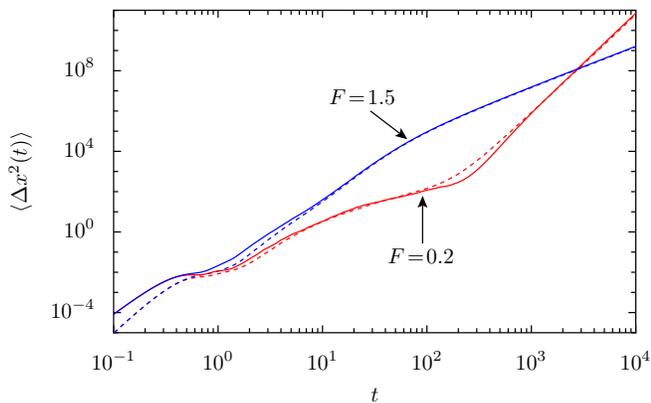}
	\caption{(Color online) Role of deviation from  stationary fluctuation-dissipation relation in Eqs. (\ref{FDT}, \ref{eqn:kubo}) on the time evolution of the mean square displacement $\langle \Delta x^2 (t)\rangle$.
	In contrast to the choice with a stationary fluctuation-dissipation relation, see solid lines marking $\langle u_i(0)u_j(0)\rangle\!=\!\kappa^2\delta_{i,j}$, the initial choice $u_1(0)\!=\!u_2(0)\!=\!0$, see dashed lines, yields a Gaussian noise $\zeta(t)$ that  initially is non-stationary, see Eq. (\ref{eqn:statnoise}).
	The noise $\zeta (t)$  assumes, however, stationary noise at asymptotic long times. 
	The initial velocity was set to zero, i.e. $v(0)=0$, the potential strength to $V_0=1$ and the remaining parameters are chosen as in Fig. \ref{Fig5},  $\kappa=2$, $\nu=3$ and $\w_{0}=1$.}
\label{Fig6}
\end{figure}

We next comment on the ergodic properties of the velocity fluctuations
in relation to
ballistic diffusion. 
In the case of free ballistic
diffusion, the velocity fluctuations are clearly non-ergodic. As already
mentioned above, this rigorously follows from the fact that the
velocity fluctuation correlation coefficient $K_{v}(t)$ given by
Eq.~(\ref{vel-corr}) converges to a
constant value different from zero. Also the strong dependence of the
position mean square displacement on the initial distribution of the
auxiliary variables $u_{1}$ and $u_{2}$ in the limit of large times, see Fig.~\ref{Fig1}, 
provides a clear indication of non-ergodicity.
In the case of ballistic diffusion in a tilted periodic potential
analytic results for the velocity fluctuation autocorrelations are
not available and we therefore have to rely on our numerical findings.
In Fig.~\ref{Fig5} the
mean square deviations of position for different
distributions of the initial velocities are compared with each other.
Apart from minor deviations,
these initial preparations do not seem to have any
influence in the presence of a tilted periodic potential. Therefore, one might suppose that in this case the process of the velocity fluctuations is wide
sense ergodic. This
result though cannot be considered conclusive because
also in the absence of any potential the choice of the
initial distribution of velocities has only little impact on the
mean square
displacement, see lines labeled by $V_0=0,F=0$ in Fig. \ref{Fig5}.
 A more convincing argument results from the comparison
of the effect of different initial distributions of the auxiliary
variables
$u_{1}$ and $u_{2}$, see Fig.~\ref{Fig6}. While the influence of this
distribution on the position mean square deviation is very large and
even increases with growing time, see Fig.~\ref{Fig1}, only small
deviations at early and intermediate times are visible in the case of
a tilted periodic potential. Hence, numerical evidence seems to
indicate that the velocity fluctuations of a ballistically diffusing
particle in a tilted washboard potential indeed is wide sense ergodic.

This raises the question whether it is possible that the velocity
fluctuations of a superdiffusive process may be wide sense ergodic in
one case and non-ergodic in another. The strict answer to this question is
that the velocity fluctuations of any truly ballistic diffusion with
$\langle\Delta x^{2}(t)\rangle = D_{2} t^{2}$ constitute
a non-ergodic process. This follows from Eq.~(\ref{sigma}) by means of
differentiation with respect to time yielding
\begin{equation}
D_{2} = \lim_{t \to \infty}\frac{1}{t} \int_{0}^{t} dt' K_{v}(t')\;.
\end{equation}
Therefore the time average of the autocorrelation function of the velocity
fluctuations does not vanish and consequently the velocity
fluctuations are non-ergodic, see Appendix~\ref{B}.

However, one must keep in mind that the ballistic diffusion presents a
marginal case. Any increase of $\langle\Delta x^{2}(t)\rangle$ slower than
$t^{2}$, such as $t^{2-\epsilon}$ with any small, positive $\epsilon$, or $t^{2}/\ln t$ will
lead  to a vanishing time average of the
velocity fluctuation autocorrelation function by the same argument as
above. From the numerical
point of view there is always a limitation of how accurate the scaling
exponent of $\langle\Delta x^{2}(t)\rangle$ can be determined. Logarithmic
corrections are almost impossible to identify. We therefore suppose
that the observed ballistic diffusion in a tilted washboard potential
might strictly speaking be marginally sub-ballistic and the velocity
fluctuations wide sense ergodic.

\section{Summary}

In this work we considered one of the simplest models for the
superdiffusive motion of a particle described by a GLE.
It corresponds to a bi-exponential memory kernel with
zero integral. The according spectral density $J(\omega)$ of
thermal bath oscillators sets in with a cubic law. It  describes, 
for example the diffusion of an impurity in a crystal.

We considered a large family of Markovian embedding schemes, i.e. higher
dimensional Markovian processes that generate the considered
non-Markovian process upon projection onto the subspace spanned by
position and momentum of the particle. Out of the whole class we
identified a simple four-dimensional embedding that can be numerically
treated in an efficient way.

We confirm that ballistic superdiffusion is non-ergodic, which is concordant with the findings in \cite{Bao05,Bao05a}.
As a new amazing manifestation of non-ergodicity we found that a non-equilibrium initial noise preparation can change the law of diffusion, see in Fig.~\ref{Fig1}.

Further on our numerical findings indicate that the free ballistic diffusion,
being present in the absence of any potential, changes into normal
diffusion in the presence of a periodic potential. We concluded that
the process of the velocity fluctuations is non-ergodic in the absence
of a periodic potential but wide sense ergodic in the presence of a
periodic potential. Apparently, the transition to the ergodic motion
does not require a minimal potential strength.
Rather,  the time to reach the asymptotic regime of
normal diffusion diverges
with vanishing
potential strength $V_{0}$.

An additional biasing force leads to ballistic motion in a periodic
potential, i.e. both the mean value and the variance of the position
displacement grow proportional to a $t^{2}$ law. In this case, however
we found strong indications that the velocity fluctuations remain wide
sense ergodic. This paradoxically looking scenario -- non-ergodic for
free ballistic diffusion versus ergodic for ballistic diffusion in a
potential -- is possible because the ballistic diffusion presents a
marginal situation. Although for ballistic diffusion following a
strict $t^{2}$ law
the velocity fluctuations are non-ergodic any modification of the
$t^{2}$ law with a weakly decaying function such as $1/\ln t$ leads to
wide sense ergodic velocity fluctuations.
For a sub-critical bias $F<F_0=2 \pi V_{0}/x_{0}$ the ballistic
diffusion coefficient $D_{2}$ is substantially enhanced
compared to the diffusion coefficient for free ballistic
diffusion. This effect is the analog to giant enhancement of
normal diffusion in tilted washboard potentials.

Depending on the potential and the bias strengths the time before the
asymptotic ballistic motion sets in may be extremely large. Within this
long transient period a normal and even a hyper-diffusion regime may
exist,where $\alpha_{\rm eff}$ exceeds the ballistic value of 2. The
presence of such long transients presents a general feature of the
studied non-Markovian dynamics.

{\bf Acknowledgements}
This work was supported by the German Excellence Initiative via the
{Nanosystems Initiative Munich} (NIM).

\appendix
\section{Conditions for embedding}\label{app:condemb}

The solution of the last equation in Eq. (\ref{eqn:inan}) is
\begin{align}
\v u (t)=&- \int_0^t e^{-\M{A} (t-t')}p(t')\v r dt' + \notag\\
&\int_0^t e^{-\M{A} (t-t')}\M{C} \v \xi(t')dt' + e^{-\M{A} t}\v u(0)\;,
\end{align}
which  inserted into (\ref{eqn:inan}) yields
\begin{align}\label{eqn:pan}
\dot p(t) =& -\left[\frac{\partial}{\partial x} V(x,t)-\v g^\mathrm{T} \v u(t)\right]= \notag\\
	=& -\frac{\partial}{\partial x} V(x,t) -\int_0^t \v g^\mathrm{T} e^{-\M{A} (t-t')}\v r p(t')dt' + \notag\\
	& \int_0^t \v g^\mathrm{T} e^{-\M{A} (t-t')}\M{C} \v \xi(t')dt' +
\v g^\mathrm{T}e^{-\M{A} t}\v u(0) \;.
\end{align}
The comparison of (\ref{eqn:pan}) with (\ref{eqn:init}) gives Eq. (\ref{first}) and
\begin{eqnarray}\label{eqn:z}
\zeta(t) = \int_0^t \v g^\mathrm{T} e^{-\M{A} (t-t')}\M{C} \v \xi(t')dt' + \v g^\mathrm{T}e^{-\M{A} t}\v u(0)\;.
\end{eqnarray}
Assuming $\langle u_i(0)\xi_j(t)\rangle = 0$, this enables one to calculate the noise correlation function:
\begin{align}\label{eqn:ncf}
& \langle \zeta(t)\zeta(s)\rangle = \langle \v g^\mathrm{T}e^{-\M{A} t}\v u(0) \v g^\mathrm{T}e^{-\M{A} s}\v u(0) \rangle \nonumber \\
& +\langle \int_0^t dt' \int_0^s ds' \v g^\mathrm{T} e^{-\M{A}(t-t')}\M{C} \v \xi(t')
 \v g^\mathrm{T} e^{-\M{A}(s-s')}\M{C} \v \xi(s')\rangle
\end{align}
Taking into account Eq. (\ref{xicorr}) for $t>s$ (the case $t<s$ can be treated alike) the second term reduces to:
\begin{align}\label{eqn:kubo12}
\int_0^s \v g^\mathrm{T} e^{-\M{A}(t-s')}\M{C}\M{C}^\mathrm{T} e^{-\M{A}^\mathrm{T}(s-s')}\v g ds'=  \notag\\
=\v g^\mathrm{T} e^{-\M{A} t}\int_0^s e^{\M{A} s'}\M{C}\M{C}^\mathrm{T} e^{\M{A}^\mathrm{T} s'}ds' e^{-\M{A}^\mathrm{T} s}\v g
\end{align}
and the first term is:
\begin{align}\label{eqn:kubo2}
&\m{\v g^\mathrm{T}e^{-\M{A} t}\v u(0) \v g^\mathrm{T}e^{-\M{A} s}\v u(0)}=  \notag\\
=&\v g^\mathrm{T}e^{-\M{A} t}\m{\v u(0)\otimes \v u^\mathrm{T}(0)} e^{-\M{A}^\mathrm{T} s}\v g \;.
\end{align}
Altogether, with the following definition:
\begin{equation}
\M{U}\equiv\m{\v u(0)\otimes \v u^\mathrm{T}(0)}
\label{U}
\end{equation}
this yields:
\begin{align}
\m{\zeta(t)\zeta(s)}=\v g^\mathrm{T} e^{-\M{A} t}\left[\M{U} + \int_0^s e^{\M{A} s'}\M{C}\M{C}^\mathrm{T} e^{\M{A}^\mathrm{T} s'}ds' \right]e^{-\M{A}^\mathrm{T} s}\v g \notag \;.
\end{align}
Making an Ansatz as in Eq.(\ref{eqn:cond}) enables one to separate the noise correlation function into a stationary and a non-stationary part:
\begin{align}
\m{\zeta(t)\zeta(s)}=&\v g^\mathrm{T} e^{-\M{A} t}\left[\M{U} + \int_0^s e^{\M{A} s'}(\M{A}\M{G}+\M{G}\M{A}^\mathrm{T}) e^{\M{A}^\mathrm{T} s'}ds' \right]\notag\\
&\times e^{-\M{A}^\mathrm{T} s}\v g =  \notag\\
=&\v g^\mathrm{T} e^{-\M{A} t}\left[\M{U} + e^{\M{A} s'}\M{G} e^{\M{A}^\mathrm{T} s'}\Big|_{s'=0}^{s'=s}
\right]e^{-\M{A}^\mathrm{T} s}\v g =  \notag\\
=&\v g^\mathrm{T} e^{-\M{A} t}\left[\M{U}-\M{G}\right]e^{-\M{A}^\mathrm{T} s}\v g + \v g^\mathrm{T} e^{-\M{A} (t-s)}\M{G} \v g
\label{eqn:statnoise}
\end{align}
The first term of Eq. (\ref{eqn:statnoise}) represents the non-stationary 
part and is vanishing asymptotically in the limit of long times, 
i.e. $t,s\to\infty$. Both the relaxation spectrum defining the
corresponding time scales and the spectrum
of autocorrelation times is given by the eigenvalues of matrix $\M{A}$. 
Moreover, the fluctuation-dissipation relation of Eq.~(\ref{FDT}) 
is always asymptotically fulfilled, if one chooses $\M{G}\v g =mk_BT\v r$, which yields  Eq. (\ref{Gtor}).
However, in order to obey the fluctuation-dissipation relation for all times, one 
has to set $\M{U}\equiv\M{G}$, which implies Eq. (\ref{eqn:UG})
 and we end with:
\begin{align}
  \m{\zeta(t)\zeta(s)}=m k_B T\v g^\mathrm{T} e^{-\M{A}(t-s)}\v r=mk_BT\gamma(t-s)\;.
  \label{eqn:kubo}
\end{align}
\section{Wide sense ergodicity}\label{B}
According to its definition, a stationary process $y(t)$ is ergodic in
the wide sense if its time average
converges in the mean square sense towards the ensemble average
. This definition implies that a process $y(t)$ is wide sense
ergodic if and only if the
time average of the autocorrelation function of its fluctuations
vanishes, i.e. if
\begin{equation}
\lim_{t \to \infty} \frac{1}{t} \int_{0}^{t} dt' \langle (y(t)-
\langle y \rangle)(y(0)-
\langle y \rangle)  \rangle = 0
\end{equation}
holds \cite{Yaglom}.
Hence, the decay of the autocorrelation function of the
fluctuations towards zero provides a sufficient condition for a wide
sense ergodic process. On the other hand, the considered process is
non-ergodic if the autocorrelation function of its fluctuations
approaches a constant value different from zero.



\begin{thebibliography}{99}
\bibitem{Jordan}If the matrix $\M{A}$ can only be brought into Jordan
  normal form, $\v g \exp (-\M{A}t) \v r$ will have
less than $N$ exponentials $\exp (-\lambda_{i} t)$ and additional
terms of the form   $t^{k} \exp (-\lambda_{i} t)$.
\end{thebibliography}
\end{document}